

\input phyzzx.tex
\input myphyx.tex
\overfullrule0pt


\def\etal{{\it et al.}}
\def\half{{\textstyle{1 \over 2}}}
\def\third{{\textstyle{1 \over 3}}}
\def\quarter{{\textstyle{1 \over 4}}}

\def\bold#1{\setbox0=\hbox{$#1$}%
     \kern-.025em\copy0\kern-\wd0
     \kern.05em\copy0\kern-\wd0
     \kern-.025em\raise.0433em\box0 }
\Pubnum={VAND-TH-93-1}
\date={January 1993}
\pubtype{}
\titlepage

\vskip1cm
\title{\bf Electroweak Corrections to the Supersymmetric Charged Higgs
Fermionic Decay}
\author{Marco Aurelio D\'\i az }
\vskip .1in
\centerline{Department of Physics and Astronomy}
\centerline{Vanderbilt University, Nashville, TN 37235}
\vskip .2in

\centerline{\bf Abstract}
\vskip .1in

One-loop radiative corrections to the charged Higgs
coupling to two fermions and to the charged Higgs fermionic decay
are studied. A
renormalization scheme is proposed including the definition of the
parameter $\tan\beta$ through the coupling of the CP-odd Higgs boson
to a charged lepton pair. Loops including top and bottom
quarks and squarks are considered using the unitary gauge.

Large corrections to the coupling are found when there is a substantial
mixing in the squark mass matrix, but these corrections are suprress
in the decay rate. Interesting applications to
rare processes are discussed.

\vfill

\endpage

\noindent{\bf 1. Introduction}
\vskip 0.5cm

\REF\guide{J.F. Gunion, H.E. Haber, G. Kane and S. Dawson,
{\it The Higgs Hunter's Guide} (Addison-Wesley, Reading MA, 1990).}
\REF\diazhaberi{M.A. D\'\i az and H.E. Haber, {\it Phys. Rev. D}
{\bf 45}, 4246 (1992).}
\REF\GtCprB{J.F. Gunion and A. Turski, {\it Phys. Rev. D} {\bf 39}, 2701
(1989);{\bf 40}, 2333 (1989); P.H. Chankowski, S. Pokorski and J. Rosiek,
{\it Phys. Lett. B} {\bf 274}, 191 (1992); A. Brignole, {\it Phys. Lett.
B} {\bf 277}, 313 (1992).}
\REF\expmch{UA2 Collaboration, J. Alitti \etal, {\it Phys. Lett. B}
{\bf 280},
137 (1992); ALEPH Collaboration, D. Decamp \etal, {\it Phys. Rept.}
{\bf 216}, 216 (1992); UA1 Collaboration, C. Albajar \etal, {\it Phys.
Lett. B} {\bf 257}, 459 (1991); VENUS Collaboration, T. Yuzuki \etal,
{\it Phys. Lett. B} {\bf 267}, 309 (1991); DELPHI Collaboration, P. Abreu
\etal, {\it Phys. Lett. B} {\bf 241}, 449 (1990); L3 Collaboration, B.
Adeba \etal, {\it Phys. Lett. B} {\bf 252}, 511 (1990); OPAL Collaboration,
M. Akrawy \etal, {\it Phys. Lett. B} {\bf 242}, 299 (1990).}
\REF\KunsztZ{Z. Kunszt and F. Zwirner, report no. CERN-TH-6150/91,
1991 (unpublished).}
\REF\QCDtbH{C.S. Li and T.C. Yuan, {\it Phys. Rev. D} {\bf 42}, 3088 (1990);
C.S. Li, Y.S. Wei and J.M. Yang, {\it Phys. Lett. B} {\bf 285}, 137 (1992);
J. Liu and Y.-P. Yao, {\it Phys. Rev. D} {\bf 46}, 5196 (1992);
A. Czarnecki and S. Davidson, Report No. Alberta-THY-34-92, December 1992,
unpublished.}
\REF\QCDHtb{A. Mendez and A. Pomarol, {\it Phys. Lett. B} {\bf 252},
461 (1990); C.S. Li and R.J. Oakes, {\it Phys. Rev. D} {\bf 43},
855 (1991).}
\REF\MPomarol{A. Mendez and A. Pomarol, {\it Phys. Lett. B}
{\bf 265}, 177 (1991).}
The Minimal Supersymmetric Model (MSSM)
has five physical states in the Higgs sector: two CP-even
neutral Higgs ($H$,$h$), one CP-odd neutral Higgs ($A$) and a
charged Higgs pair ($H^{\pm}$)\refmark\guide. Radiative corrections to
the charged Higgs mass have been found to be substantial in the region
of small or very large values of $\tan\beta$ (the ratio between the
vacuum expectation values of the two Higgs doublets) and small
CP-odd Higgs mass\refmark{\diazhaberi,\GtCprB}. When there is no mixing
in the squark mass matrix, these corrections are proportional to the
square of the top quark mass. Nevertheless, when squark
mixing is present the radiative corrections turn out to be proportional
to $m_t^4$. The search for the charged Higgs boson
at LEP has set a lower bound on its mass given by $m_{H^{\pm}}>$
41.7 GeV, where it is assumed that the dominant decays are $H^+$ to
$\tau^+\nu_{\tau}$ and $c\bar s$\refmark\expmch.
Many of the properties of the
charged Higgs (if it is discovered) will be explored at the LHC and
the SSC, where for low values of the charged Higgs mass the dominant
production mechanism is $gg\rightarrow t\bar t$ followed by
$t\rightarrow H^+b$. The dominant decay modes for $m_{H^{\pm}}<m_t$ are
$\tau^+\nu_{\tau}$ and $c\bar s$. Previous calculations\refmark\KunsztZ
include radiative corrections to the charged Higgs mass
and leading QCD corrections to the vertices\refmark{\KunsztZ,\QCDtbH,
\QCDHtb}.
The purpose of this paper
is to calculate electroweak radiative corrections to the $H^+$ vertex
to two fermions (excluding the $tb$ case)
in the context of the MSSM. These corrections were
found to be small in a non-supersymmetric two-Higgs doublet
model\refmark\MPomarol, but as we will see, the inclusion of squarks is
important especially when a large mixing is present in their mass
matrix.

\REF\diazhaberii{M.A. D\'\i az and H.E. Haber, {\it Phys. Rev. D} {\bf 46},
3086 (1992).}
We start by a renormalization prescription of the parameter
$\tan\beta$ proposed first in reference [\diazhaberi].
That is followed by the renormalization of the $H^+$ vertex to two
fermions. We work in the unitary gauge where the Goldstone bosons
are decoupled, and we consider one-loop involving only top/bottom
quark/squarks. The contributions due to gauge bosons, higgses, neutralinos
and charginos are in general smaller (see for example ref. [\diazhaberii]).

\vskip 0.3cm
\noindent{\bf 2. Renormalization scheme}
\vskip 0.5cm

\REF\Hollik{W.F.L. Hollik, {\it Fortsch. Phys.} {\bf 38}, 165 (1990).}
The counterterms are defined
through the relations between the bare quantities at the left and
the renormalized quantities at the right\refmark\Hollik:
$$\eqalign{&\Psi_L\longrightarrow(Z_L)^{1\over 2}\Psi_L=
(1+{\textstyle{1\over 2}}\delta Z_L)\Psi_L,\qquad\quad \Psi_L=
{u\choose d}_L...{\nu_e\choose e}_L...\cr
&\psi_R\longrightarrow(Z_R)^{1\over 2}\psi_R=(1+{\textstyle{1\over
2}}\delta Z_R)\psi_R,\qquad\quad\psi_R=u_R,d_R...,e_R...\cr
&H_i\longrightarrow(Z_{H_i})^{1\over 2}H_i=(1+{\textstyle{1\over 2}}
\delta Z_{H_i})H_i,\qquad\quad i=1,2\cr
&W_{\mu}^a\longrightarrow (Z_W)^{1\over 2}W_{\mu}^a=(1+{\textstyle
{1\over 2}}\delta Z_W)W_{\mu}^a,\qquad a=1,2,3\cr
&B_{\mu}\longrightarrow(Z_B)^{1\over 2}B_{\mu}=(1+{\textstyle
{1\over 2}}\delta Z_B)B_{\mu}\cr}\eqn\ecuxiii$$
for the fields, and
$$\eqalign{
&\lambda\longrightarrow \lambda-\delta\lambda,\qquad\qquad\qquad\qquad
\qquad\quad\lambda=g,g',e...\cr
&m\longrightarrow m-\delta m,\qquad\qquad\qquad\qquad\qquad
m=m_W...,m_{12},...\cr
&v_i\longrightarrow v_i-\delta v_i,\qquad\qquad\qquad\qquad\qquad
\quad i=1,2.\cr}\eqn\ecuxiv$$
for the coupling constants, the masses and
the vacuum expectation values.

The renormalization of the parameter $\tan\beta$ is done through
the $Al^+l^-$ coupling. The decay rate of a CP-odd Higgs into a
charged lepton pair is:
$$\Gamma(A\longrightarrow l^+l^-)={{g^2m_l^2}\over{32\pi m_W^2}}
m_A\tan^2\beta\eqn\ecuxviii$$
and this equation relates the parameter $\tan\beta$ with an observable
given by the experimental value of the rate. Of course, since there is no
experimental evidence for the CP-odd Higgs, eq.~\ecuxviii\ relates two
unknown parameters ($\tan\beta$ and the rate). Therefore we will treat
$\tan\beta$ as an input parameter.
The renormalized vertex is
$$\Gamma^{Al^+l^-}(p^2)=-{{gm_lt_{\beta}}\over{2m_W}}
\bigl[1+\Lambda^{Al^+l^-}(p^2)-{{\delta g}\over
g}-{{\delta m_l}\over{m_l}}-{{\delta t_{\beta}}\over{t_{\beta}}}
+{{\delta m_W}\over{m_W}}+{\textstyle{1\over 2}}\delta Z_A+
\delta Z_V^l\bigr]\gamma_5,\eqn\ecui$$
where $\Lambda^{Al^+l^-}(p^2)$ are the one-loop contributions
to the vertex. The wave function renormalization constants of the
CP-odd Higgs and the charged lepton, are defined by:
$$\delta Z_A=s_{\beta}^2\delta Z_{H_1}+c_{\beta}^2\delta Z_{H_2}
\qquad\qquad\delta Z_V^l=\half(\delta Z_L^{\nu l}+\delta Z_R^l)
\eqn\ecuxv$$
The one-loop contributions to the vertex come from the mixing between
the $Z$ gauge boson and the CP-odd Higgs: denoting by $A_{AZ}(p^2)
p^{\mu}$ the Feynman diagrams contributing to the $A-Z$ two point
function, we have:
$$\Lambda^{Al^+l^-}(p^2)=-{1\over{t_{\beta}}}{{A_{AZ}(p^2)}\over
{m_Z}}.\eqn\ecuii$$
Identifying the tree level vertex with the renormalized one at the
scale $p^2=m_A^2$, we solve for the $\tan\beta$ counterterm:
$${{\delta t_{\beta}}\over{t_{\beta}}}=-\Lambda^{Al^+l^-}(m_A^2)
-{{\delta g}\over g}-{{\delta m_l}\over{m_l}}+{{\delta m_W^2}
\over{2m_W^2}}+{\textstyle{1\over 2}}\delta Z_A+\delta Z^l_V.
\eqn\ecuiii$$
Our normalization for the gauge boson masses is such that
$m_W^2=\quarter g^2v^2$ and $m_Z^2=\quarter(g^2+{g'}^2)v^2$,
where $v^2=v_1^2+v_2^2$. Using these relations and the definition of
the electric charge $e=gs_W=g'c_W$, we find the following expression
for $\delta g$:
$${{\delta g}\over g}={{\delta e}\over e}-{{c_W^2}\over{
2s_W^2}}\Bigl({{\delta m_Z^2}\over{m_Z^2}}-{{\delta m_W^2}\over
{m_W^2}}\Bigr).\eqn\ecuiv$$
In an on-shell scheme, where the electric charge is defined through
the coupling $\gamma e\bar e$ at zero momentum and where there is no
mixing between the photon and the $Z$ gauge boson at zero
momentum, the counterterm $\delta e$ is:
$${{\delta e}\over e}=\Lambda_V^{\gamma e\bar e}(0)+{\textstyle{
1\over 2}}\delta Z_{\gamma}+\delta Z_V^e-{{1-4s_W^2}\over{4s_Wc_W}}
{{A_{\gamma Z}(0)}\over{m_Z^2}},\eqn\ecuv$$
where $\Lambda_V^{\gamma e\bar e}(p^2)$ are the one-loop contributions
to the $\gamma e\bar e$ vertex that are proportional to $\gamma^{\mu}$:
$\Gamma^{\gamma e\bar e}(p^2)=ie[\gamma^{\mu}+\Lambda_V^{\gamma
e\bar e}(p^2)\gamma^{\mu}+...]$.
The wave function renormalization constants of the photon and the
CP-odd Higgs are obtained fixing the residue of each propagator to
one:
$$\delta Z_{\gamma}\equiv s_W^2\delta Z_W+c_W^2\delta Z_B=A'_{\gamma
\gamma}(0),\qquad\qquad \delta Z_A=A'_{AA}(m_A^2)\eqn\ecuxvi$$
and the gauge boson mass counterterms are given by the self energies:
denoting by $iA_{ab}(p^2)g^{\mu\nu}+iB_{ab}(p^2)p^{\mu}
p^{\nu}$ with $ab=WW,ZZ,\gamma\gamma,Z\gamma$, the Feynman diagrams
that contribute to the two point functions, we have:
$$\delta m_W^2={\rm Re}A_{WW}(m_W^2),\qquad\delta m_Z^2=
{\rm Re}A_{ZZ}(m_Z^2).\eqn\ecuxvii$$
With eqs.~\ecuiv\ to~\ecuxvii\ we can solve for the $\tan\beta$
counterterm given in eq.~\ecuiii. Note that because we are considering
loops involving only third generation of quarks and squarks, $\delta
m_l$, $\delta Z_V^l$ and $\Lambda_V^{\gamma e\bar e}$ give zero
contribution. Also, although the sum of the
contributions of top and bottom quarks
and squarks to $A_{\gamma Z}(p^2)$ is not zero for $p^2\neq 0$,
they cancel exactly when we evaluate at $p^2=0$.

\vskip 0.3cm
\noindent{\bf 3. Renormalization of the {\bf $H^+d\bar u$} vertex}
\vskip 0.5cm

The renormalized $H^+d\bar u$ vertex is given by:
$$\eqalign{\Gamma^{H^+d\bar u}(p^2)=&
i\lambda_{H^+d\bar u}^-(1-\gamma_5)\left[1+\Lambda^{H^+d\bar u}_-(p^2)
-{{\delta\lambda_{H^+d\bar u}^-}\over{\lambda_{H^+d\bar u}^-}}
+\half\delta Z_{H^{\pm}}+\delta Z_V^d\right]\cr
+&i\lambda_{H^+d\bar u}^+(1+\gamma_5)\left[1+\Lambda^{H^+d\bar u}_+(p^2)
-{{\delta\lambda_{H^+d\bar u}^+}\over{\lambda_{H^+d\bar u}^+}}
+\half\delta Z_{H^{\pm}}+\delta Z_V^u\right],\cr}\eqn\ecuvi$$
where the wave function renormalization constants for the quarks are
$\delta Z_V^d=\half(\delta Z_L^{ud}+\delta Z_R^d)$ and $\delta Z_V^u=
\half(\delta Z_L^{ud}+\delta Z_R^u)$, the wave function renormalization
constant of the charged Higgs is $\delta Z_{H^{\pm}}=s_{\beta}^2
\delta Z_{H_1}+c_{\beta}^2\delta Z_{H_2}$,
the $\lambda^{\pm}$ couplings are defined by:
$$\lambda_{H^+d\bar u}^-={{gm_u}\over{2\sqrt{2}m_Wt_{\beta}}},
\qquad \lambda_{H^+d\bar u}^+={{gm_dt_{\beta}}\over{2\sqrt{2}
m_W}},\eqn\ecuvii$$
and the counterterms are:
$${{\delta\lambda_{H^+d\bar u}^-}\over{\lambda_{H^+d\bar u}^-}}
={{\delta g}\over g}+{{\delta m_u}\over{m_u}}-{{\delta
m_W^2}\over{2m_W^2}}-{{\delta t_{\beta}}\over{t_{\beta}}},
\qquad {{\delta\lambda_{H^+d\bar u}^+}\over{\lambda_{H^+d\bar u}^+}}
={{\delta g}\over g}+{{\delta m_d}\over{m_d}}-{{\delta
m_W^2}\over{2m_W^2}}+{{\delta t_{\beta}}\over{t_{\beta}}}.
\eqn\ecuviii$$
The $\Lambda_{\pm}^{H^+d\bar u}(p^2)$ are the one-loop contributions
to the vertex $H^+d\bar u$ and, in our approximation, they come from
the mixing
between the $W$ gauge boson and the charged Higgs. Denoting by
$iA_{H^+W^-}(p^2)p^{\mu}$ the Feynman diagrams contributing to the
$H^+-W^-$ two point function, we have:
$$\Lambda_-^{H^+d\bar u}(p^2)=t_{\beta}{{A_{H^+W^-}(p^2)}\over{
m_W}},\qquad \Lambda_+^{H^+d\bar u}(p^2)=-{1\over{t_{\beta}}}{{
A_{H^+W^-}(p^2)}\over{m_W}}.\eqn\ecuix$$

In the approximation where we consider only top/bottom quarks/squarks,
the renormalized $H^+d\bar u$ vertex is:
$$\Gamma^{H^+d\bar u}(p^2)=i\lambda^-_{H^+d\bar u}(1-\gamma_5)f^-(p^2)
+i\lambda^+_{H^+d\bar u}(1+\gamma_5)f^+(p^2),\eqn\ecux$$
where the $f^{\pm}$ factors are:
$$\eqalign{f^-(p^2)=&1+
\half A'_{H^+H^-}(p^2)+\half A'_{AA}(m_A^2)-A'_{\gamma\gamma}(0)
+t_{\beta}{{A_{H^+W^-}(p^2)}\over{m_W}}\cr
-&{{A_{AZ}(m_A^2)}\over{t_{\beta}m_Z}}
-{{c_{2W}}\over{s_W^2}}{{A_{WW}(m_W^2)}\over{m_W^2}}
+{{c_W^2}\over{s_W^2}}{{A_{ZZ}(m_Z^2)}\over{m_Z^2}},\cr
f^+(p^2)=&1+\half A'_{H^+H^-}(p^2)-\half A'_{AA}(m_A^2)
-{{A_{H^+W^-}(p^2)}\over{t_{\beta}m_W}}+{{A_{AZ}(m_A^2)}\over{
t_{\beta}m_Z}}.\cr}\eqn\ecuxi$$
and $p^2$ is the four-momentum of the charged Higgs, equal to the
renormalized charged Higgs mass if it is on-shell.
But since we are renormalizing the Green function, $H^{\pm}$
may be off-shell. In this case we need to modify eq.~\ecuxi\ with
the replacement:
$$A'_{H^+H^-}(p^2)\longrightarrow {{A_{H^+H^-}(p^2)-A_{H^+H^-}(
m_{H^{\pm}}^2)}\over{p^2-m_{H^{\pm}}^2}}\eqn\offshell$$
and $m_{H^{\pm}}$ is the renormalized charged Higgs mass.

We are using the $H^+d\bar u$ vertex notation, but the results are valid
also for $s\bar c$, $e\bar\nu_e$, $\mu\bar\nu_{\mu}$ and $\tau\bar\nu_{
\tau}$. In the case of leptons $f^-$ is irrelevant because $\lambda
^-_{H^+l\bar\nu_l}=0$ since it is proportional to the neutrino mass.
Note that the tree level value for the factors $f^{\pm}$ is one, as we see
fron eq.~\ecuxi.

Since we want the radiatively corrected
decay widths of the charged Higgs to a pair of fermions, we
need to evaluate the momentum $p^2$ of the $H^+$ leg at the renormalized
charged Higgs mass\refmark\diazhaberi:
$$p^2=m_{H^{\pm}}^2=m_W^2+m_A^2+A_{H^+H^-}(m_W^2+m_A^2)-
A_{WW}(m_W^2)-A_{AA}(m_A^2).\eqn\ecuxii$$

Each term in eq.~\ecuxi\ is infinite by itself, nevertheless, the factors
$f^{\pm}$ are measurable quantities and they must be finite. This provides
a very useful mechanism to check the calculations.

\vskip 0.3cm
\noindent{\bf 4. Results}
\vskip 0.5cm

It is useful to obtain an asymptotic formula for $f^{\pm}$ by evaluating
the exact one-loop results in the limit $M_Q^2\gg m_t^2\gg m_W^2$ and
assuming $m_A^2={\cal O}(m_W^2)$. The result for $f^+$ is
$$f^+(m^2_{H^{\pm}})\approx 1+{{N_cg^2}\over{32\pi^2m_W^2}}\biggl[
{{m_t^2}\over{4t_{\beta}^2}}+
m_b^2(1+\half t_{\beta}^2)\log{{m_t^2}\over{m_A^2}}+m_b^2(
{\textstyle{3\over 2}}+\quarter t_{\beta}^2)\biggr]\eqn\fplus$$
and for $f^-$ we have
$$\eqalign{
f^-(m_{H^{\pm}}^2)&\approx\ 1+{{N_cg^2}\over{32\pi^2m_W^2}}\Biggl\{
\half m_t^2\biggl[-{{c_W^2}\over{s_W^2}}+\third(r_3^2-r_1^2)-{1\over{
2t_{\beta}^2}}-\third r_1r_3{{c_{2\beta}}\over{s_{\beta}c_{\beta}}}
\biggr]-{{m_A^2}\over{3s_{\beta}^2}}\cr
&-\third m_W^2\biggl[{\textstyle{1\over 9}}+{1\over{2s_W^2}}+{\textstyle
{40\over 27}}s_W^2+{1\over{t_{\beta}^2}}-\bigl({\textstyle{4\over 3}}
-{1\over{s_W^2}}\bigr)\ln{{m_t^2}\over{m_Z^2}}-{\textstyle{8\over 9}}
s_W^2\ln{{m_Z^2}\over{m_b^2}}\biggr]\cr
&+m_b^2\biggl[-1-8{{c_W^2}\over{s_W^2}}(\quarter+e_bs_W^2+2e_b^2s_W^4)
+{3\over{2s_W^2}}+{1\over{t_{\beta}^2}}+{\textstyle{3\over 4}}t_{
\beta}^2+{\textstyle{1\over 6}}(r_4^2-r_2^2)\cr
&\qquad\qquad +\bigl(-2+{1\over{s_W^2}}-\half t_{\beta}^2\bigr)
\ln{{m_t^2}\over{m_Z^2}}+(1+\half t_{\beta}^2)\ln{{m_A^2}\over{
m_Z^2}}\biggr]\Biggr\}
\cr}\eqn\fminus$$
where the $r_i$ are defined in terms of the various squark mixing
parameters:
$$r_1={{\mu+A_U/t_{\beta}}\over{M_Q}},\quad r_2={{\mu+A_Dt_{\beta}}
\over{M_Q}},\quad r_3={{\mu/t_{\beta}-A_U}\over{M_Q}},\quad
r_4={{\mu t_{\beta}-A_D}\over{M_Q}}.\eqn\defri$$

The $f^-$ asymptotic formula is evaluated numerically and compared
with the ``exact'' one-loop answer given by eq.~\ecuxi, where by
``exact'' we mean the formulae given in the appendix. The comparison is
shown in Fig. 1, where we see that the asymptotic formula for $f^-$ is
very good for values of $m_A$ less than 100 GeV, as is expected since
the formulas are valid for $m_A\ll m_t$. The tree level values of $f^{
\pm}$ are unity, and we see that the radiative corrections to $f^-$
are of the order 5 to 20 \% over the whole range of $m_A$, for the
values of $\tan\beta$ chosen. The effect of the discontinuity
of the slope of the function $Re B_0(m_{H^{\pm}}^2;m_t^2,m_b^2)$ that
appears in the wave function renormalization of the charged
Higgs, is apparent at $m_A\approx$ 130 GeV. The asymptotic formula
starts to deviate from the ``exact'' one for values of $m_A$ of the order
of $m_t$ or bigger. The comparison of the ``exact'' and asymptotic
formulas for $f^+$ are equally good. However, the radiative corrections
to $f^+$ are smaller.

\REF\BargerHP{V. Barger, J.L. Hewett and R.J.N. Phillips, {\it Phys. Rev. D}
{\bf 41}, 3421 (1990).}
In Fig. 2 it is displayed the dependence of the factors $f^{\pm}$
as a function of $\tan\beta$. The lower limit $\tan\beta=0.25$ is chosen
because it is the boundary of the
region where the perturbative analysis is
valid: for lower values of $\tan\beta$, the $tbH^{\pm}$ coupling
becomes too large. For any value of $m_t$, the constraint we are using
to ensure perturbatively small couplings
is: $\tan\beta\ge m_t/$(600 GeV)\refmark\BargerHP.
Radiative corrections to $f^+$ (Fig. 2a) are small, {\it i.e.}
$f^+$ is close to one, and they correspond
basically to the difference between our scheme, and the scheme
of ref. [\MPomarol], where the radiative corrections to $f^+$
are defined to be zero. In the two
cases where $m_A$ is small, the curves are stopped at low values of
$\tan\beta$ where the radiatively corrected charged
Higgs mass reaches the lower bound of 41.7 GeV.

It was said before that the
$f^{\pm}$ must be finite because they are measurable quantities,
nevertheless, we just saw that they are scheme dependent. To clarify
what we mean, it must be said that the decay rate $\Gamma(H^+
\rightarrow u\bar d)$ is the quantity that must be scheme independent.
This decay rate depends on $f^{\pm}$ and $\tan\beta$ (as we will see
below), therefore, going from our scheme to the scheme in
ref.~[\MPomarol]\ implies a change in the values of $\tan\beta$ and
$f^{\pm}$ in such a way that the rate is unchanged.

Unlike $f^+$, radiative corrections to $f^-$ (Fig. 2b) can be
very important
when there is a substantial mixing in the squark mass matrix and the
parameter $\tan\beta$ is large. This effect can be understood by looking
at the asymptotic formula for $f^-$ given in eq.~\fminus: the term
proportional to $m_t^2r_1r_3c_{2\beta}/(s_{\beta}c_{\beta})$ is absent
in $f^+$, is zero when there is no squark mixing (or more precisely,
it is zero when $r_1$ or $r_3$ are zero), and becomes very large
when $\tan\beta\gg 1$. It is interesting to notice that $r_1r_3
\rightarrow -\mu A_U/M_Q^2$ when $\tan\beta\rightarrow\infty$, which means
that the term we are analyzing is negative when $\mu$ and $A_U$ have
the same sign (the case of Fig. 2), and it is positive when they have
the opposite sign, making $f^->1$. This effect can be seen in Fig. 3,
where we plot $f^-$ as a function of the top quark mass for different
choices of the mixing in the squark mass matrix. When there is
no mixing ($\mu=A_U=A_D=0$) radiative corrections are small and $f^-$
decreases with $m_t$, as is confirmed by the asymptotic formula in
eq.~\fminus. And if we have the same (different) relative sign between
$\mu$ and the $A$ parameters, $f^-$ decreases (increases) with $m_t$.
The bulge near $m_t\approx 130$ GeV is due to the change of slope
of the function $Re B_0(m_{H^{\pm}}^2;m_t^2,m_b^2)$ at the point
where the top quark mass is close to the corrected charged Higgs mass.

The decay rate of a charged Higgs into a quark pair is:
$$\Gamma(H^+\longrightarrow c\bar s)={{N_cg^2}\over{32\pi m_W^2}}
m_{H^{\pm}}\bigl[(f^-)^2m_c^2\cot^2\beta+(f^+)^2m_s^2\tan^2\beta\bigr]
\eqn\ecuxix$$
and to a pair of leptons:
$$\Gamma(H^+\longrightarrow \nu_{\tau}\tau^+)={{g^2}\over
{32\pi m_W^2}}m_{H^{\pm}}(f^+)^2m_{\tau}^2\tan^2\beta\eqn\ecuxx$$
These decay rates are plotted in Fig. 4 for two choices of the mass of
the CP-odd Higgs: (a) $m_A=100$ GeV and (b) $m_A=250$ GeV. In the case
of the leptonic decay the radiative corrections are dominated by
the factor $f^+$; therefore they are small except for $\tan\beta<1$ and
$\tan\beta\gg 1$. In the case of the decay to two quarks, radiative
corrections are dominated by $f^+$ for $\tan^2\beta\gsim m_c/m_s$ and
by $f^-$ for $\tan^2\beta\lsim m_c/m_s$. However, the large corrections
to $f^-$ at large values of $\tan\beta$ are suppressed in the partial width
$\Gamma(H^+\longrightarrow c\bar s)$ due to the factor $\cot^2\beta$.

\vglue 0.3cm
\noindent{\bf 5. Concluding Remarks}
\vglue 0.5cm

Large corrections to the charged Higgs coupling to two fermions can
be found if there is a substantial mixing in the squark mass matrix.
The biggest correction occurs when $\tan\beta$ is large, but non-neglegible
corrections appear also if $\tan\beta$ is smaller than the unity.
In the case of the decay rate, the corrections at large $\tan\beta$ are
suppressed by a factor of $\cot^2\beta$, with only the effects
at $\tan\beta<1$ remaining significant.

\REF\diazbtosf{M.A. D\'\i az, Report No. VAND-TH-93-2, January 1993,
unpublished.}
Radiative corrections to the charged-Higgs-fermion-fermion vertex may
have important applications to rare decay calculations, for example
correcting the one-loop induced decay $b\rightarrow s\gamma$
\refmark\diazbtosf. The reason is as follows: a one-loop graph with
a charged Higgs in the loop that couples at least twice to a pair of
fermions, may have a term proportional to $\lambda^+\lambda^-$,
with the $\lambda$ couplings defined in eq.~\ecuvii. This term will be
renormalized by a factor of $f^+f^-$, and large corrections to $f^-$
will not be supressed since $\cot\beta$ in $\lambda^-$ is canceled with
$\tan\beta$ in $\lambda^+$. This is the case in the one-loop induced
decay $b\rightarrow s\gamma$, and may also be the case for other rare
processes.

\vglue 0.5cm
{\bf \noindent Acknowledgements }
\vglue 0.4cm
Discussions with Howard Haber, Ralf Hempfling, Alex Pomarol
and Thomas Weiler are
gratefully acknowledged. This work was supported in part by the U.S.
Department of Energy.

\REF\PVeltman{G. Passarino and M. Veltman, {\it Nucl. Phys.}
{\bf B160}, 151 (1979).}

\refout

\FIG\funo{Comparison between the ``exact'' one-loop formula for $f^-$
and its asymptotic expansion given by eq.~\fminus\ as a function of
$m_A$, for two different
values of $\tan\beta$: 1.0 and 30. All the supersymetry breaking mass
parameters in the squark sector are taken to be equal to $M_{SUSY}$
(this is valid for all the figures).}

\FIG\fdos{Radiatively corrected factor (a) $f^+$ and (b) $f^-$ as a
function of $\tan\beta$ at $p^2=m_{H^{\pm}}^2$.
The $f^+$ ($f^-$) factor renormalizes the term proportional to the
down(up)-type fermion mass in the $H^+d\bar u$ vertex.}

\FIG\ftres{Radiatively corrected factor $f^-$ as a function of $m_t$ for
different choices of the mixing in the squark mass matrix.}

\FIG\fcuatro{Influence of the one-loop radiative corrections to the charged
Higgs decay rate into a pair of fermions as a function of $\tan\beta$.
Two values of $m_A$ are considered: (a) 100 GeV and (b) 250 GeV. We
contrast the tree level answer with the radiatively corrected one, for
the decay products $\tau^+\nu_{\tau}$ and $c \bar s$.}

\figout

\endpage
\Appendix {A}

In this appendix we display the exact one-loop formulae we need
to compute the radiatively corrected charged Higgs coupling to a
pair of fermions, for loops involving top and bottom quarks and
squarks. We follow the notation of ref. [\PVeltman] for the loop
integrals, up to a change in the metric:
$$\eqalign{
&A_0(m^2)=m^2(\Delta-\ln\ m^2+1),\cr
&B_0(p^2;m_1^2,m_2^2)=\Delta-\int_0^1dx\ln\ [m_2^2x+m_1^2(1-x)
-p^2x(1-x)-i\epsilon],\cr
&B_{22}(p^2;m_1^2,m_2^2)=\quarter(\Delta+1)(m_1^2+m_2^2-\third p^2)-\cr
&\half\int_0^1dx[m_2^2x+m_1^2(1-x)-p^2x(1-x)]\ln\ [m_2^2x
+m_1^2(1-x)-p^2x(1-x)-i\epsilon],\cr
&B_1(p^2;m_1^2,m_2^2)={1\over{2p^2}}[A_0(m_1^2)-A_0(m_2^2)-
(p^2+m_1^2-m_2^2)B_0(p^2;m_1^2,m_2^2)],\cr
}\eqn\loopint$$
where $\Delta$ is the regulator of dimensional regularization defined
by:
$$\Delta={2\over{4-n}}+\ln\ 4\pi-\gamma_E,\eqn\deltadef$$
$n$ is the number of space-time dimensions, and $\gamma_E$ is
Euler's constant.
The formulas displayed below are written explicitly for the third
generation of quarks and squarks, but can be easily relabeled for
any other weak doublet of fermions and supersymmetric partners.
First we give the quark contribution identifying it with a superindex
$tb$, and then the squark contribution with a superindex $\tilde t
\tilde b$.
We start with the photon wave function renormalization constant
$\delta Z_{\gamma}=A'_{\gamma\gamma}(0)$:
$$\Big[A'_{\gamma\gamma}(0)\Big]^{tb}=-{{N_cg^2s_W^2}\over{8\pi^2}}
\bigg[e_t^2(4{B'_{22}}^{0tt}+{B_0}^{0tt})+e_b^2(4{B'_{22}}^{0bb}+
{B_0}^{0bb})\bigg]\eqn\dfotfottb$$
and:
$$\Big[A'_{\gamma\gamma}(0)\Big]^{\tilde t\tilde b}=
{{N_cg^2s_W^2}\over{4\pi^2}}\bigg[e_t^2({B'_{22}}^{0\tilde t_1
\tilde t_1}+{B'_{22}}^{0\tilde t_2\tilde t_2})+e_b^2({B'_{22}}
^{0\tilde b_1\tilde b_1}+{B'_{22}}^{0\tilde b_2\tilde b_2})
\bigg]\eqn\dfotfotstb$$
where $N_c=3$ is the number of colors, $e_t$ and $e_b$ are the
electric charges of the top and bottom quarks, $B_{22}^{0xy}=
B_{22}(0;m_x^2,m_y^2)$ and
$B'_{22}$ is the derivative of $B_{22}$
with respect to the first argument.

The derivative of the charged Higgs self energy with respect to
the momentum $p^2$ is:
$$\eqalign{
\bigl[A'_{H^+H^-}(p^2)\bigr]^{tb}&={{N_cg^2}\over{32\pi^2m_W^2}}
\biggl\{\cr
(m_t^2/t_{\beta}^2&+m_b^2t_{\beta}^2)\Bigl[
-B_0^{ptb}+(m_t^2+m_b^2-p^2){B'_0}^{ptb}\Bigr]+
4m_t^2m_b^2{B'_0}^{ptb}\biggr\}\cr}\eqn\ahhtb$$
and
$$\eqalign{
&\bigl[A'_{H^+H^-}(p^2)\bigr]^{\tilde t\tilde b}=-{{N_cg^2}\over{
32\pi^2m_W^2}}\biggl\{\cr
&\Bigl[{{m_t^2}\over{t_{\beta}^2}}+m_b^2t_{\beta}^2-m_W^2s_{2\beta}
\Bigr]^2(c_t^2c_b^2{B'_0}^{11}+c_t^2s_b^2{B'_0}^{12}+s_t^2c_t^2
{B'_0}^{21}+s_t^2s_b^2{B'_0}^{22})\cr
&+{{m_t^2m_b^2}\over{s_{\beta}^2c_{\beta}^2}}(s_t^2s_b^2{B'_0}^{11}+
s_t^2c_b^2{B'_0}^{12}+c_t^2s_b^2{B'_0}^{21}+c_t^2c_b^2{B'_0}^{22})\cr
&+2s_tc_ts_bc_b{{m_tm_b}\over{s_{\beta}c_{\beta}}}\Bigl[{{m_t^2}\over
{t_{\beta}^2}}+m_b^2t_{\beta}^2-m_W^2s_{2\beta}\Bigr]({B'_0}^{11}-
{B'_0}^{12}-{B'_0}^{21}+{B'_0}^{22})\cr
&+2s_tc_tm_tr_1M_Q\Bigl[{{m_t^2}\over{t_{\beta}^2}}+m_b^2t_{\beta}^2
-m_W^2s_{2\beta}\Bigr](c_b^2{B'_0}^{11}+s_b^2{B'_0}^{12}-c_b^2{B'_0}^
{21}-s_b^2{B'_0}^{22})\cr
&+2s_bc_bm_br_2M_Q\Bigl[{{m_t^2}\over{t_{\beta}^2}}+m_b^2t_{\beta}^2
-m_W^2s_{2\beta}\Bigr](c_t^2{B'_0}^{11}-c_t^2{B'_0}^{12}+s_t^2{B'_0}^
{21}-s_t^2{B'_0}^{22})\cr
&+2s_bc_b{{m_t^2m_b}\over{s_{\beta}c_{\beta}}}r_1M_Q(s_t^2{B'_0}^{11}
-s_t^2{B'_0}^{12}+c_t^2{B'_0}^{21}-c_t^2{B'_0}^{22})\cr
&+2s_tc_t{{m_tm_b^2}\over{s_{\beta}c_{\beta}}}r_2M_Q(s_b^2{B'_0}^{11}
+c_b^2{B'_0}^{12}-s_b^2{B'_0}^{21}-c_b^2{B'_0}^{22})\cr
&+m_t^2r_1^2M_Q^2(s_t^2c_b^2{B'_0}^{11}+s_t^2s_b^2{B'_0}^{12}+c_t^2
c_b^2{B'_0}^{21}+c_t^2s_b^2{B'_0}^{22})\cr
&+m_b^2r_2^2M_Q^2(c_t^2s_b^2{B'_0}^{11}+c_t^2c_b^2{B'_0}^{12}+s_t^2
s_b^2{B'_0}^{21}+s_t^2c_b^2{B'_0}^{22})\cr
&+2s_tc_ts_bc_bm_tm_br_1r_2M_Q^2({B'_0}^{11}-{B'_0}^{12}-{B'_0}^{21}
+{B'_0}^{22})\biggr\}
\cr}\eqn\ahhstb$$
where $B_0^{pxy}=B_0(p^2;m_x^2,m_y^2)$, ${B'_0}^{ij}$ is the derivative
of $B_0(p^2;m_{\tilde t_i}^2,m_{\tilde b_j})$ with respect to the
first argument, and $s_t$, $c_t$, $s_b$ and $c_b$ are trigonometric
functions of the squark mixing angles.

The CP-odd wave function renormalization constant
is given by $\delta Z_A=A'_{AA}(m_A^2)$ where:
$$\left[A'_{AA}(m_A^2)\right]^{tb}=-{{N_cg^2}\over{32\pi^2m_W^2}}\left[
{{m_t^2}\over{t_{\beta}^2}}(B_0^{Att}+m_A^2{B'_0}^{Att})+
m_b^2t_{\beta}^2(B_0^{Abb}+m_A^2{B'_0}^{Abb})\right]\eqn\dAAtb$$
and:
$$\left[A'_{AA}(m_A^2)\right]^{\tilde t\tilde b}=-{{N_cg^2}\over{
32\pi^2m_W^2}}\left[m_t^2(\mu+A_U/t_{\beta})^2{B'_0}^{A\tilde
t_1\tilde t_2}+m_b^2(\mu+A_Dt_{\beta})^2{B'_0}^{A\tilde b_1
\tilde b_2}\right]\eqn\dAAstb$$
here, $B_0^{Axy}=B_0(m_A^2;m_x^2,m_y^2)$.

The $Z$ gauge boson self energy is given by:
$$\eqalign{\Big[A_{ZZ}(m_Z^2)\Big]^{tb}=&
{{N_cg^2}\over{32\pi^2c_W^2}}(m_t^2B_0^{Ztt}+m_b^2B_0^{Zbb})\cr
-&{{N_cg^2}\over{16\pi^2c_W^2}}(\quarter-e_ts_W^2+2e_t^2s_W^4)
(4B_{22}^{Ztt}-2A_0^t+m_Z^2B_0^{Ztt})\cr
-&{{N_cg^2}\over{16\pi^2c_W^2}}(\quarter+e_bs_W^2+2e_b^2s_W^4)
(4B_{22}^{Zbb}-2A_0^b+m_Z^2B_0^{Zbb})\cr}\eqn\ZZtb$$
where $B_{22}^{Ztt}\equiv B_{22}(m_Z^2;m_t^2,m_t^2)$, and:
$$\eqalign{\Big[A_{ZZ}&(m_Z^2)\Big]^{\tilde t\tilde b}=
{{N_cg^2}\over{8\pi^2c_W^2}}\Big[\cr
&(-\half c_t^2+e_ts_W^2)^2(2B_{22}^{Z\tilde t_1\tilde t_1}
-A_0^{\tilde t_1})+
(\half c_b^2+e_bs_W^2)^2(2B_{22}^{Z\tilde b_1\tilde b_1}
-A_0^{\tilde b_1})+\cr
&(-\half s_t^2+e_ts_W^2)^2(2B_{22}^{Z\tilde t_2\tilde t_2}
-A_0^{\tilde t_2})+
(\half s_b^2+e_bs_W^2)^2(2B_{22}^{Z\tilde b_2\tilde b_2}
-A_0^{\tilde b_2})+\cr
&\quarter s_t^2c_t^2(4B_{22}^{Z\tilde t_1
\tilde t_2}-A_0^{\tilde t_1}-A_0^{\tilde t_2})+
\quarter s_b^2c_b^2(4B_{22}^{Z\tilde b_1
\tilde b_2}-A_0^{\tilde b_1}-A_0^{\tilde b_2})\Big].\cr}
\eqn\ZZstb$$
where $A_0^x=A_0(m_x^2)$.

The $W$ gauge boson self energy is:
$$\Big[A_{WW}(m_W^2)\Big]^{tb}=-{{N_cg^2}\over{32\pi^2}}\biggl[
4B_{22}^{Wtb}-A_0^t-A_0^b+(m_W^2-m_t^2-m_b^2)B_0^{Wtb}\biggr]
\eqn\WWtb$$

$$\eqalign{\Big[A_{WW}(m_W^2)\Big]^{\tilde t\tilde b}=&
{{N_cg^2}\over{8\pi^2}}\Big[c_t^2c_b^2B_{22}^{W\tilde t_1
\tilde b_1}+c_t^2s_b^2B_{22}^{W\tilde t_1\tilde b_2}+s_t^2c_b^2
B_{22}^{W\tilde t_2\tilde b_1}+s_t^2s_b^2B_{22}^{W\tilde t_2
\tilde b_2}\Big]\cr
&-{{N_cg^2}\over{32\pi^2}}\Big[c_t^2A_0^{\tilde t_1}+s_t^2
A_0^{\tilde t_2}+c_b^2A_0^{\tilde b_1}+s_b^2A_0^{\tilde b_2}
\Big]\cr}\eqn\WWstb$$

The mixing between the charged Higgs and the $W$ gauge boson is
given by:
$$\Big[A_{H^+W^-}(p^2)\Big]^{tb}={{N_cg^2}\over{16\pi^2m_W}}
\left[{{m_t^2}\over{t_{\beta}}}B_1(p^2;m_b^2,m_t^2)-m_b^2
t_{\beta}B_1(p^2;m_t^2,m_b^2)\right]\eqn\AHWtb$$

$$\eqalign{\Big[A_{H^+W^-}&(p^2)\Big]^{\tilde t\tilde b}=
{{N_cg^2}\over{32\pi^2m_W}}\bigg\{\cr
&(m_t^2/t_{\beta}+m_b^2t_{\beta}-m_W^2s_{2\beta})\Big[c_t^2c_b^2B^{11}+
c_t^2s_b^2B^{12}+s_t^2c_b^2B^{21}+s_t^2s_b^2B^{22}\Big]+\cr
&m_b(\mu+A_Dt_{\beta})s_bc_b\Big[
c_t^2B^{11}-c_t^2B^{12}+s_t^2B^{21}-s_t^2B^{22}\Big]+\cr
&m_t(\mu+A_U/t_{\beta})s_tc_t\Big[
c_b^2B^{11}+s_b^2B^{12}-c_b^2B^{21}-s_b^2B^{22}\Big]+\cr
&{{m_tm_b}\over{s_{\beta}c_{\beta}}}s_tc_ts_bc_b\Big[
B^{11}-B^{12}-B^{21}+B^{22}\Big]
\bigg\}\cr}\eqn\AHWstb$$
where $B^{ij}\equiv 2B_1(p^2;m_{\tilde t_i}^2,m_{\tilde b_j}^2)
+B_0(p^2;m_{\tilde t_i}^2,m_{\tilde b_j}^2)$.

Finally, the $A$-$Z$ mixing is:
$$\Big[A_{AZ}(p^2)\Big]^{tb}=-{{N_cg^2}\over{32\pi^2c_Wm_W}}
\bigg[{{m_t^2}\over{t_{\beta}}}B_0(p^2;m_t^2,m_t^2)-m_b^2
t_{\beta}B_0(p^2;m_b^2,m_b^2)\bigg]\eqn\AZtb$$
and:
$$\eqalign{
\Big[A_{AZ}(p^2)\Big]^{\tilde t\tilde b}={{N_cg^2}\over{
32\pi^2c_Wm_W}}\bigg[&m_t(\mu+A_U/t_{\beta})s_tc_t(B_1^{
p\tilde t_1\tilde t_2}-B_1^{p\tilde t_2\tilde t_1})\cr
-&m_b(\mu+A_Dt_{\beta})s_bc_b(B_1^{p\tilde b_1\tilde b_2}
-B_1^{p\tilde b_2\tilde b_1})\bigg]\cr}\eqn\AZstb$$
\end